\documentclass{ws-p8-50x6-00}

\def\ifmath#1{\relax\ifmmode #1\else $#1$\fi}%
\def\rd{\ifmath{{\mathrm{d}}}}
\def\rt{\ifmath{{\mathrm{t}}}}

\def\QCD{\ifmath{{\mathrm{QCD}}}}
\def\tag{\ifmath{{\mathrm{tag}}}}

\def\true{\ifmath{{\mathrm{true}}}}
\def\jets{\ifmath{{\mathrm{jets}}}}
\def\hadrons{\ifmath{{\mathrm{hadrons}}}}
\def\vis{\ifmath{{\mathrm{vis}}}}

\begin{document}

\title{Tests of QCD in two photon physics at LEP}

\author{A.J. Finch}

\address{Lancaster University, Lancaster, LA1 4YW, United Kingdom\\
E-mail: A.Finch@lancaster.ac.uk}


\maketitle

\abstracts{
Recent developments are reviewed in the field of two photon physics at
LEP and their contribution to testing QCD.  }

\section{Introduction}
\begin{figure}[t]
%\figurebox{20pc}{15pc}{} % to have a box alone
\epsfysize=5.3cm % will enlarge or reduce the postscript figures based on the xsize
\begin{centering}
\epsfbox{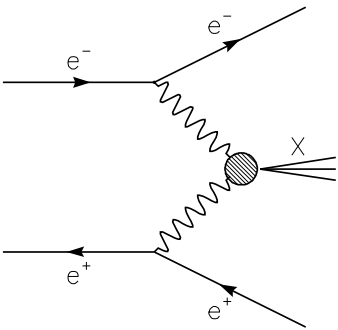} % postscript image file name
\caption{Photon photon collisions as observed at an e$^+$e$^-$ collider
  \label{diagram}}
\end{centering}
\end{figure}

Two photon physics is performed at e$^+$e$^-$ colliders such as LEP via
the interaction of two virtual photons, as shown in
Fig.~\ref{diagram}. The energy spectrum of the virtual photons is
sharply peaked at low energy, so that the final states produced are
predominantly at low energy. However, the cross section for the process 
rises logarithmically with the e$^+$e$^-$ centre of mass
energy, so that at LEP II, with no $\mathrm{Z}^0$ resonance to compete with, it
is the dominant process. The following standard variables are defined.
\begin{list}{$\bullet$}{\setlength{\itemsep}{-0.1em}}
\item
$Q^2$ = -(4 momentum $\mathrm{transfer)^2}$ of photon 1
\item
$P^2$ = -(4 momentum $\mathrm{transfer)^2}$ of photon 2
\item[]\hspace{0.5cm} $Q^2 > P^2$ by definition
\item
$W$ = Invariant mass of final state
\item
 $x=\frac{Q^2}{Q^2+W^2}$
\end{list}

A number of facts make the experimental measurements in
two photon physics difficult. The first of these is that
one does not have a monoenergetic beam of photons. Instead,
all measurements must be extracted from a measured e$^+$e$^-$
process. Secondly, in general the observed invariant mass
$W_{\vis}$ is less than the true one $W_{\true}$. This
is due to the tendency for much of the final state energy either 
to be deposited in
the forward luminosity spectrometers, which are poor at measuring
hadronic energy, or even be lost along the beam pipe.

From the theoretical point of view, the photon, as a fundamental
particle of the Standard Model, should allow precise calculations to be
performed. This is indeed the case, provided it is probed with
sufficiently large a scale that perturbative QCD
calculations can be performed. However, many of the photons in two
photon collisions have small four momentum, which can not be described
by perturbative QCD. In these cases, we are forced to adopt a
phenomenological model, Vector Meson Dominance, where the photon is
assumed to behave like a vector meson.  This means that, in order for
measurement made at LEP to test QCD, they have to be limited to
processes which include at least one large scale. The scales available
are
\begin{list}{$\bullet$}{\setlength{\itemsep}{-0.1em}}
\item
Four momentum of photon(s)
\item
Transverse momentum of final state
\item
Mass of produced quark
\item
Mass of  produced hadron pair.
\end{list}

The following sections  discuss recent advances in
measurements using the first three of these. Most of these results
were reported at the PHOTON 2000 conference,\cite{PHOTON2000} the 
proceedings of which will be published by AIP in 2001. Prior to 
publication, they can be seen at
{\sf http://www.photon2000.lancs.ac.uk/proceedings/}.
\section{The Photon Structure Function}

 The photon structure function $F_2^\gamma(x,Q^2)$ is measured
directly in so-called single tagged events, in which one of the
electrons is scattered sufficiently to be detected. This results in
the $Q^2$ of the associated photon being significantly
different from zero. In this arrangement, it is then acceptable to
distinguish this photon as a probe of  the properties of the
other photon, whose four momentum transfer will be close to zero, and which
is treated as the target. This in turn means that the full machinery of deep
inelastic scattering can be adopted. The cross section for
the process is expressed   in terms of two structure functions
$F_1^\gamma(x,Q^2)$ and $F_2^\gamma(x,Q^2)$,

\[\frac{\rd\sigma}{\rd x\rd y}=\frac{4\pi\alpha^2s(e\gamma)}{Q^4}[(1-y)
F_2^\gamma(x,Q^2)+xy^2F_1^\gamma(x,Q^2)\ , \]

where

 \[y=1-\frac{E_{\tag}}{E}\cos^2(\frac{\theta_{\tag}}{2})\ . \]
However, as $y$ is very   much less than one in any practical
measurement, only $F_2^\gamma$ is eperimentally accessible.

When  $F_2^\gamma(x,Q^2)$ is calculated, it falls naturally into two parts,
one part  known as the point-like part is calculable exactly in QCD
and depends only on $\lambda_{\QCD}$. The second, so-called hadronic
part is not calculable and has to be approximated by some flavour of
VDM. A number of different authors have attempted to calculate
 $F_2^\gamma(x,Q^2)$, resulting in sets of parton distribution functions
(PDFs) which can be compared to data.
% are known  as usual by various acronyms. The three most commonly used
%in comparison to data are GRV (Gl\"{u}ck, Reya and Vogt)~\cite{GRV},
%SAS (Schuler and Sj\"{o}strand)~\cite{SAS}
%and LAC (Levy, Abramowicz, and Charchula)~\cite{LAC}. As well as being tested
%directly in photon structure function measurements, these different
%PDFs are used in calculating other processes in two photon physics
%such as high $p_t$ and heavy flavours.

 While measurements of the Photon Structure Function have been made
since the late 1970's, there continue to be improvements at LEP II. In
addition to much larger data sets, higher $Q^2$ and lower $x$
than ever before, there has been progress in two areas aimed at
reducing the large systematic errors that plague the
measurement. These arise from the problem of unfolding the true final
state invariant mass distribution from the observed one. This requires
a good model of the final state, which has been lacking for many
years, resulting in large model dependance of the result. Two
approaches have been taken to attack this. The first is the
improvement of the Monte Carlo models by the
production of standardized hadron level distributions by three of the
four LEP experiments.\cite{LEPwide} This is intended to make it easier for
 program authors to tune their programs. The other new approach
is the use of two-dimensional unfolding techniques to reduce the model
dependance of the result by using additional information that is
available in the data.  This approach has been adopted by ALEPH~\cite{ALEPHf2}
and OPAL.\cite{OPALf2}

\section{High $p_\rt$}

In the production of high transverse momentum jets, or particles, it
is the transverse momentum scale which `probes' the structure of a
photon. Calculations are performed by convoluting  matrix
elements for the hard scattering process with flux factors of photons
produced by electrons, and photon PDFs.
 Photons are classified as `Direct' if the photon enters
the hard scattering process itself or `Resolved' if it is a parton
from the photon which is involved. Events are classified as Direct,
Single Resolved or Double Resolved, depending on how many photons in
the event are resolved. Apart
from kinematical effects,  these three processes are all of the same
order of magnitude in cross section.

Very recently, 
two authors~\cite{Klasen,Poetter} have provided code to allow NLO QCD
calculations to be performed in	 a Monte Carlo manner which allows
experimental cuts to be applied to the calculation. This greatly
improves the possibilities of comparing data and theory, though it
should be noted that such comparision is still between theory
calculated with massless quarks, and data corrected to the hadron level.

Recent experimental results include the first measurement of jet
production in tagged events by a LEP experiment~\cite{ALEPHtjet}
 allowing access in principle to the
virtual photon structure function.
They agree well with the NLO predictions in $p_\rt$ distributions.
 A new approach from OPAL~\cite{OPALjet}
is an attempt to get a more direct probe of differences between the
various photon PDFs. These show greatest difference at low $x$,
however these differences are largely lost when measuring $p_\rt$
distributions. Instead,  OPAL have plotted $x_{\gamma}$ which
is defined as
$x_\gamma^\pm = \frac{ \sum_{\jets}{(E \pm p_z)}}
                     { \sum_{\hadrons}{(E \pm p_z)}}$
(both entries ($x_\gamma^+$ and $x_\gamma^-$) are included in   distributions).
Comparing to the equivalent distributions in Monte Carlo shows some
direct sensitivity to the PDF at low $x$, however comparing to NLO
calculations emphasizes the difference between calculations done with
partons and hadron level measurements, as very poor agreement is
found. The NLO calculation produces far too large a spike at $x = 1$
with correspondingly too low a cross section at lower $x$ values.

\section{Heavy Flavours}

From the theoretical point of view, the measurement of heavy flavour
production in two photon physics has much to recommend it. The only
processes that contribute significantly are the direct and single
resolved ones.\cite{DKZZ} The contribution of the latter is dominated by
processes including initial state gluons and thus directly probes the
gluonic part of the photon PDFs, which is not directly measured in
photon structure function measurements. The scale of the process is
set by the heavy quark mass. In tagged measurements the situation is
even more favourable, as the direct and resolved components do not mix
and are scheme independent up to NLO. For $x > 0.2$, the pointlike part
dominates and is exactly calculable in QCD. Below this, the hadronic part,
and hence the gluon part of the photon, is probed.\cite{Laenen}

 Performing the  measurement is made difficult by the lack of a good
high statistics tag for charm or beauty in two photon physics. The
lifetime tagging techniques which have been so succesful in
e$^+$e$^-$ annihilation events have not turned out to be useful. Two
heavy flavour tags have been used, the `$\mathrm{D}^*$
mass trick', which has the advantage  of being a clear unambiguous
charm tag, but with low statistics, and lepton tagging, which suffers 
from large poorly understood backgrounds. Early measurements of charm 
production were presented as total cross sections. These suffer from 
large errors, when extrapolating from the observed
visible cross section to the total cross section.
The results are consistent with NLO QCD predictions, but these
predictions have large mass and scale dependancies.
Recently, more differential calculations have become 
available,\cite{Kraemer} and it turns out that
these also have smaller theoretical uncertainties. As a result,
LEP measurements are now presented in limited $p_\rt$ and rapidity
corresponding to experimental acceptances. Current measurements from
L3~\cite{L3charm} and OPAL~\cite{OPALcharm}
 agree well with each other and with a massless
calculation, while ALEPH~\cite{ALEPHcharm}
 preliminary measurements show a flatter $p_\rt$
dependance, and agree better with the massive calculation.
L3 also show the charm cross section as a function of
invariant mass and find a NLO calculation reproduces the slope of
their data, but not the normalisation which is a factor of two above
the calculation.

 In lepton tagged data, both L3~\cite{L3charm} and OPAL~\cite{OPALcharm}
 observe an excess at high
$p_\rt$ which they ascribe to b-quark production. The rate observed by
each experiment is consistent but 2$\sigma$ above the theoretical expectation.

 A first measurement of tagged charm production has been presented by
OPAL.\cite{OPALcharmf2} There are two bins in $x$. The high $x$ bin is 
consistent with the NLO calculation, but the low $x$ point suffers from 
too large errors to make any meaningful comparisons.

\section{Double Tagged Measurements}

 Untill recently, the lack of statistics has meant that little
attention has been paid to double tagged events. However, the large
amount of data available at LEP means that this is now a feasible
area of study and allows to test the obervation~\cite{Balitski}
that for events with
 $W^2_{\gamma^*\gamma^*} >> (Q_1^2,Q_2^2)$
 the BFKL calculations  are considerably larger in certain areas of
phase space
than those using the DGLAP approach. This was seen as a clean way to
distinguish the two. Early results from L3~\cite{L32tag} were clearly above the
predictions of models such as PHOJET although not as large as the LO
BFKL prediction. Recently, it has become clear that NLO corrections to
BFKL are large,\cite{BFKLNLO} reducing the difference between them and DGLAP
calculations. In addition, OPAL~\cite{OPAL2tag} have shown that there
is a possibility of
large radiative corrections being required when
calculating the final invariant mass from the tagged electrons. The
corrections go in the direction of reducing the cross section just in
the kinematical region where large deviations from DGLAP
were apparently observed
by L3. It would seem premature at this stage to say whether  BFKL
calculations can really be tested in the LEP double tagged data.

\section{Conclusions}
The great success of the LEP machine at CERN has provided us with an
unrivalled sample of two photon physics events to study. The analysis of
these events remains challenging, but in recent years the greater
collaboration between the experiments and theorists
has led to much improvement in the detail to which the data can be used to
probe QCD. It is to be hoped that the end of LEP does not lead to too
rapid an end to this effort as there is still much to be learned.


\begin{thebibliography}{99}
%
\bibitem{PHOTON2000} {\em PHOTON 2000} ed. A. Finch
(American Institute of Physics, Ambleside, 2001)
\bibitem{LEPwide}ALEPH Coll. and L3 Coll. and OPAL Coll. and 
LEP Working Group, CERN-EP-2000-109, submitted to Eur. Phys. J. C.
\bibitem{ALEPHf2} A. B\"ohrer, {\em Nucl. Phys.} B (Proc. Suppl) 
{\bf 82}, {25} (2000).
\bibitem{OPALf2}OPAL Collab.,  G. Abbiendi {\em et al}, 
{\em Eur. Phys. J.} C {\bf 18}, 15 (2000).
\bibitem{Klasen}M. Klasen and G. Kramer, {\em Z. Phys.} C {\bf 76}, 67
(1997).
\bibitem{Poetter}B. P\"otter, {\em Comp. Phys. Comm.} {\bf 119}, {45} (1999).
\bibitem{ALEPHtjet}P. Hodgson in [1].
\bibitem{OPALjet}
B. Surrow and T. Wengler in [1].
\bibitem{DKZZ} M. Drees, M. Kr\"amer, J. Zunft and P.M. Zerwas, 
{\em Phys. Lett.} B {\bf 306}, {371} (1993).
\bibitem{Laenen}E. Laenen and S. Riemersma, {\em Phys. Lett.} B {\bf 376},
{169} (1996).
\bibitem{Kraemer}S. Frixione, M.Kr\"amer, and E. Laenen, 
{\em Nucl. Phys.} B {\bf 571}, {169} (2000).
\bibitem{OPALcharm}A. Csilling, in [1].
\bibitem{L3charm} S. Saremi,  in [1].
\bibitem{ALEPHcharm}U. Sieler,  in [1].
\bibitem{OPALcharmf2}OPAL, G. Abbiendi {\em et al}, (1999), hep-ex/9911030
\bibitem{Balitski}Ya.Ya. Balitski and L.N. Lipatov,
{\em Sov. J. Nucl. Phys.} {\bf 28}, {822} (1978).
\bibitem{L32tag}L3 Collab., M. Acciari {\em et al}, 
{\em Phys. Lett.} B {\bf 453}, {333} (1999).
\bibitem{BFKLNLO}V.S. Fadin and L.N. Lipatov, 
{\em Phys. Lett.} B {\bf 429}, {127} (1998).
\bibitem{OPAL2tag}M. Przybycie\'{n},  in [1].
\end{thebibliography}
\end{document}
%%%%%%%%%%%%%%%%%%%%%%%%%%%%%%%%%%%%%%%%%%%%%%%%%%%%%%%%%%%%%%%%%%%%%%%%%%%%%
%% End of ws-p8-50x6-00.tex
%%%%%%%%%%%%%%%%%%%%%%%%%%%%%%%%%%%%%%%%%%%%%%%%%%%%%%%%%%%%%%%%%%%%%%%%%%%%%